\documentclass{pramana}

\usepackage{graphicx,amsmath,bm}

\begin{document}

\title{Numerical Estimation of Primary Ionization with Noble Gases}

\author{R. Kanishka\textsuperscript{*}}
\affilOne{Department of Physics, University Institute of Sciences, Chandigarh University, Mohali, Punjab, India 140413\\}


\twocolumn[{

\maketitle

\corres{kanishka.rawat.phy@gmail.com}


\begin{abstract}
In high energy physics experiments, the numerical estimation of primary ionization is crucial. An advance study on primary ionization can help in minimizing the effects such as electrical discharges that can damage the gaseous detectors used in high energy physics experiments. The simulation of primary ionization of electrons and positrons with the noble gases using the geant4 toolkit have been presented. The obtained primaries have been separated from secondaries as the later contribute towards electrical discharges and further damage the detectors. The xenon gas shows highest primary ionization among the noble gases.
\end{abstract}



}]

\doinum{:}
\artcitid{\#\#\#\#}
\volnum{}
\year{2024}
\pgrange{1--7}
\setcounter{page}{1}
\lp{7}

\section{Introduction}
\label{intro}

The particle physics experiments like LHC-CMS \cite{citation1}--\cite{citation5}, ALICE-TPC \cite{citation6}--\cite{citation10}, LHCb \cite{citation11}--\cite{citation15}, ATLAS \cite{citation15b}--\cite{citation15c}, and INO-RPC \cite{citation16}--\cite{citation20}, comprises of gaseous ionization detectors. These gaseous detectors works on the principle of ionization. These detectors have been fabricated and characterized to work for many years. But due to the prolonged use of beam and/or radiation, they can have various negative consequences from immediate to a long-term damage to the detectors. For the long-term operation, these detectors face problems like radiation hardness, aging of detectors and, electrical discharges \cite{citation21}--\cite{citation25}. These electrical discharges creates short-circuit between the electrodes of the detectors that damage these detectors and cause them non-operational. The current work aims to study the primary ionization which is crucial for the gaseous detectors \cite{citation26} used in nuclear and high energy physics experiments. In this paper, the properties of charged particles like primary ionization, spatial distributions, separation of primaries and secondaries using the geant4 toolkit \cite{citation27} have been presented. The simulation of the charged particles i.e., electrons and positrons have been done in different noble gases \cite{citation26}. The noble gases are beneficial for gaseous detectors as they possess good dielectric properties i.e., they do not breakdown easily at high voltage which are required for gaseous ionization detectors. The paper has been organized as follows: section 2 describes the methodology, section 3 describes the results, and summary and conclusions have been discussed in the section 4.

\section{Methodology}
\label{method}
\begin{figure*}
\centering\includegraphics[height=.3\textheight]{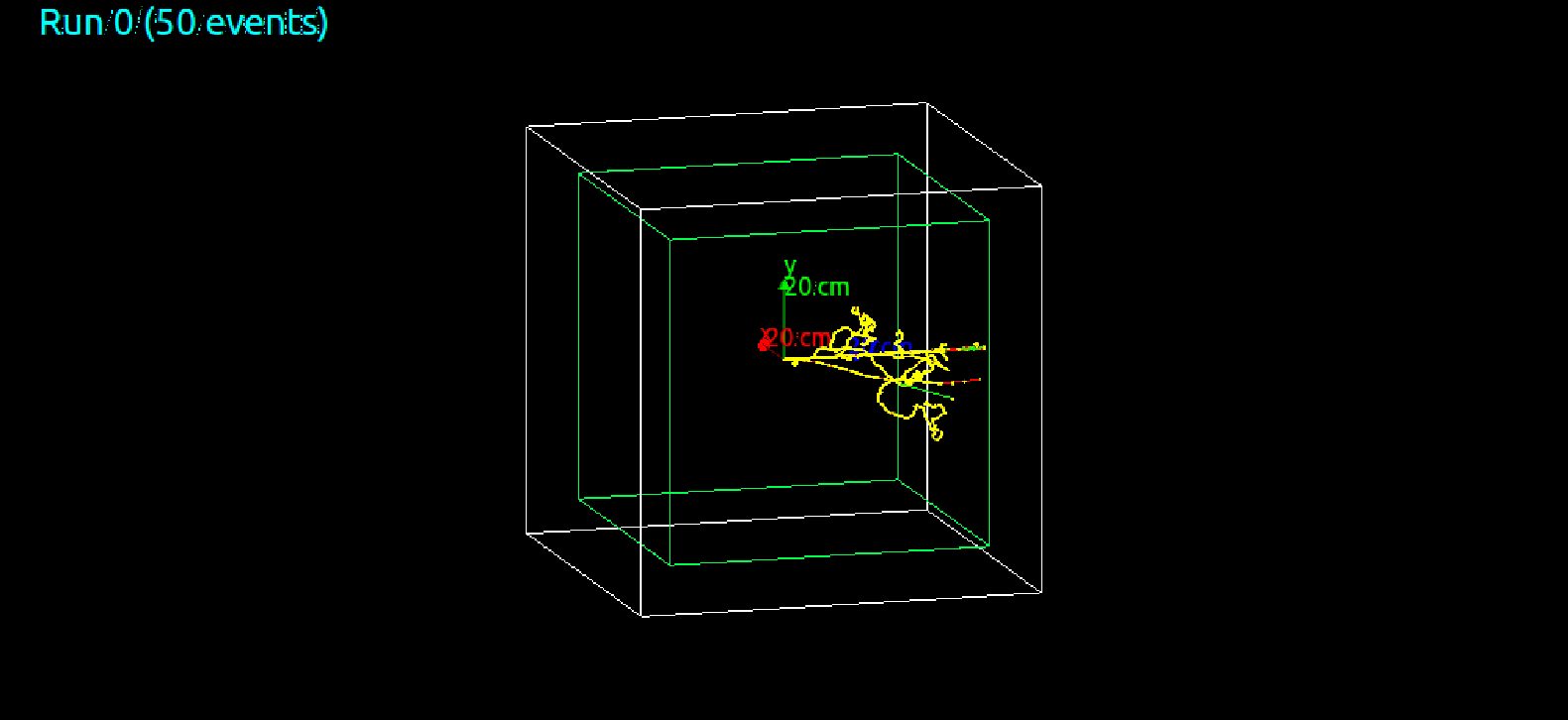}
\caption{The visualization of simulation of interaction of electrons in xenon gas. Note: The red axis shows the x-direction, green axis shows the y-direction and blue axis shows z-direction.}\label{figOne}
\end{figure*}

\begin{figure*}
\centering\includegraphics[height=.3\textheight]{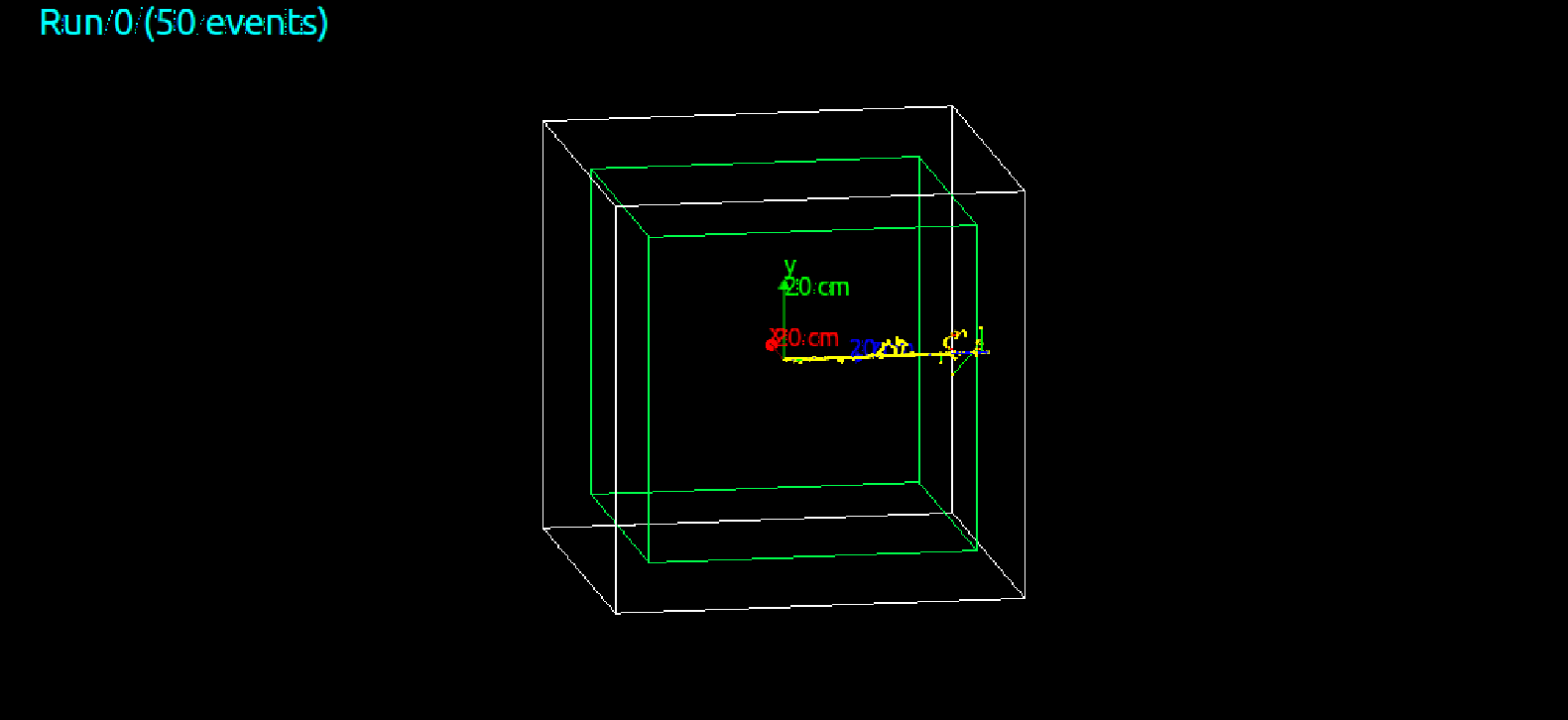}
\caption{The visualization of simulation of interaction of positrons in xenon gas. Note: The red axis shows the x-direction, green axis shows the y-direction and blue axis shows z-direction.}\label{figTwo}
\end{figure*}

The geant4 toolkit \cite{citation27} have been used to simulate electrons and positrons passing through the medium. The geant4 can simulate particle interactions at a broad range of energies, including those between electrons, photons, and heavy ions. The geant4 program's modular design enables users to customize the simulation's constituent parts like the experimental setup's shape, the materials to be used, the particle sources, and the particular kinds of physical interactions that require modeling. It's applications include several domains beyond particle physics, including medical physics, space sciences, and materials sciences. The data that has been obtained through geant4 has been analyzed with ROOT software \cite{citation28} which is an open-source framework for data analysis, used mainly for high energy physics. In our geant4 program the gas volume, physics interactions, position, direction, type of particles, and energy have been incorporated. The noble gases i.e., Helium, Neon, Argon, Krypton and Xenon have been chosen for the study. The noble gases of higher atomic number such as Radon have not been considered since that is a radioactive gas. This is beyond the scope of this paper. The steps i.e., prestep from which the particles start and poststep at which the particles udergo physical interactions in the gas volume have been defined. 10,000 particles (electrons or positrons) of 1 GeV energy inside a gas of volume of 40$\times$40$\times$40 $cm^{3}$ have been simulated. The initial position of particle was set to x=0, y=0 and z=40cm so that they can be shot along positive z-direction. The figure~\ref{figOne} and figure~\ref{figTwo} shows the geant4 display of simulation of interaction of electrons and positrons in xenon gas respectively. The simulation of 50 charged particles of 1 GeV energy that were shot in a xenon gas volume of x = y = z = 40 cm dimensions have been shown. The white colour box represents the imaginary laboratory and the detector filled with the gas volume detector have been shown in green colour. The initial position of the electron beam was set to x = y = 0 cm and z = 40 cm and shot along the z direction. The physics lists such as EMLivermore and EMPenelope have been incorporated in our study as these are used for the electromagnetic processes. The red colour beam shows electron beam and yellow coloured points shows the ionization of xenon gas. The formation of gamma particles have been shown with the green coloured lines. The results have been discussed in the next section.

  \section{Results}
\label{rs}  
The simulation was done with the electrons and positrons in different gases using geant4 which have been discussed in the next sub-sections.

\subsection{Simulation of Electrons and Positrons}
\label{el}
The different noble gases that have been mentioned in section~\ref{method} i.e., Helium, Neon, Argon, Krypton and Xenon were used. These gases acts as a target for ionization. The results obtained with electrons and positrons have been discussed below.

\begin{figure*}
\centering\includegraphics[height=.3\textheight]{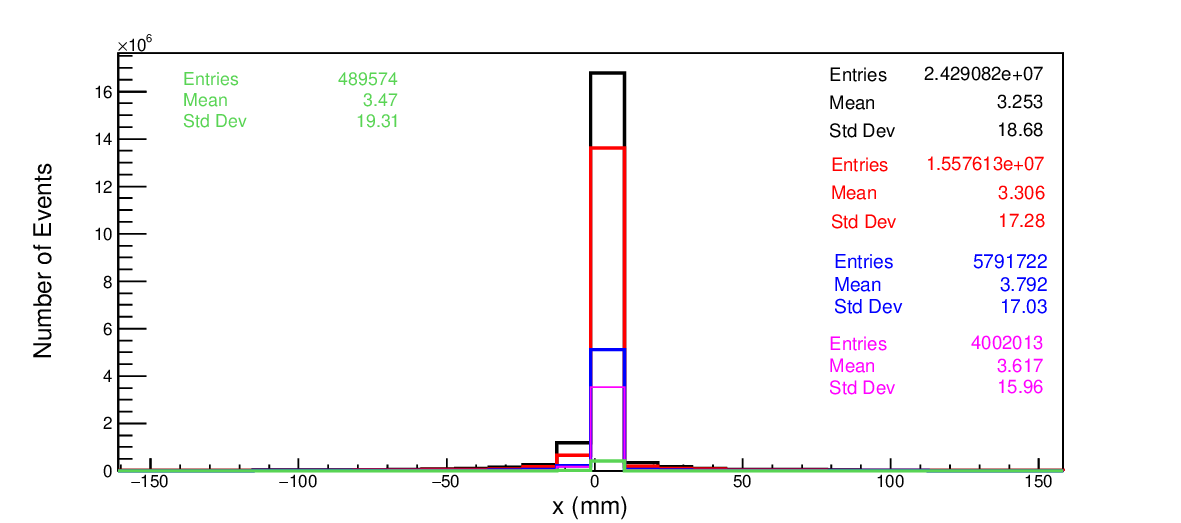}
\centering\includegraphics[height=.3\textheight]{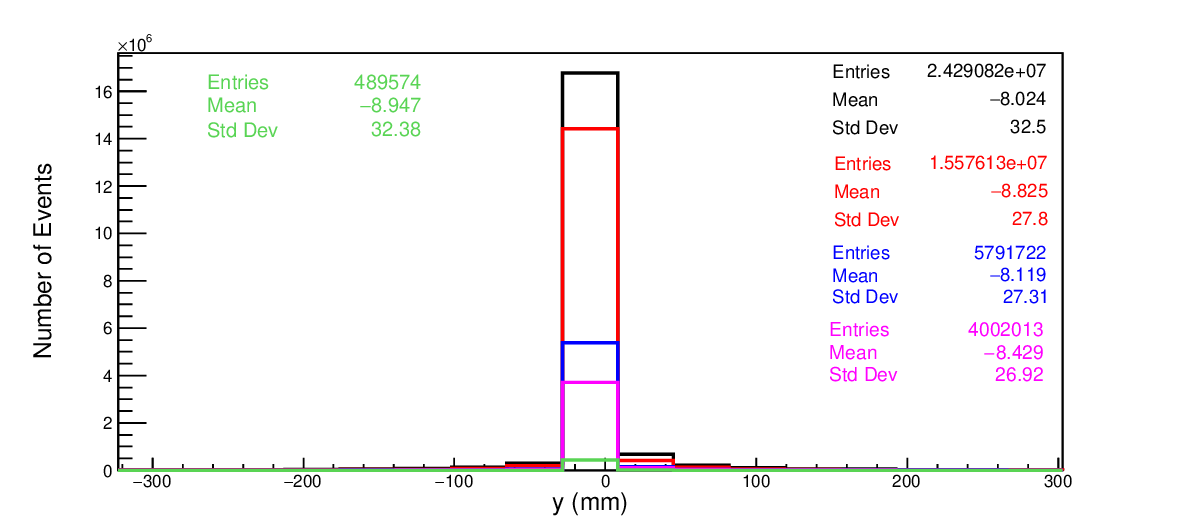}\\
\centering\includegraphics[height=.3\textheight]{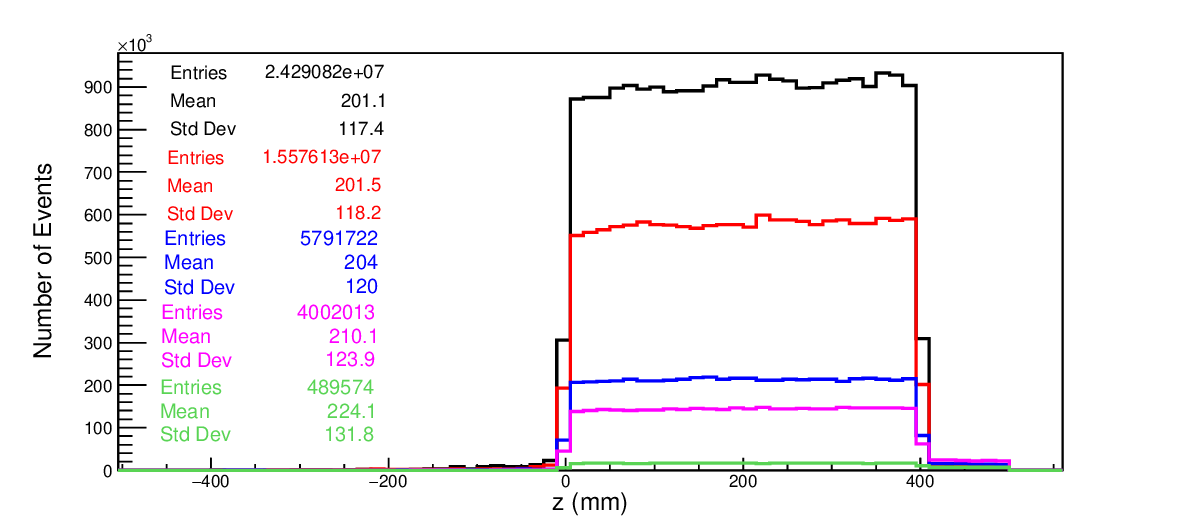}

\caption{The x, y, z distributions of the primaries obtained from simulation of electrons using five different noble gases respectively. Where black, red, blue, pink and green represents xenon, krypton, argon, neon and helium respectively.}\label{figThree}
\end{figure*}

\begin{figure*}
\centering\includegraphics[height=.3\textheight]{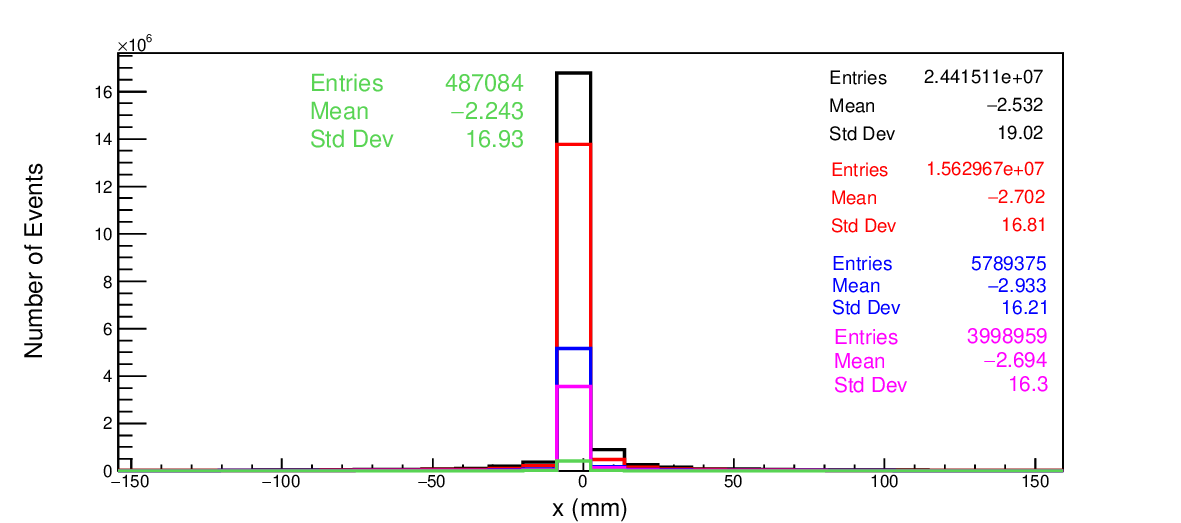}
\centering\includegraphics[height=.3\textheight]{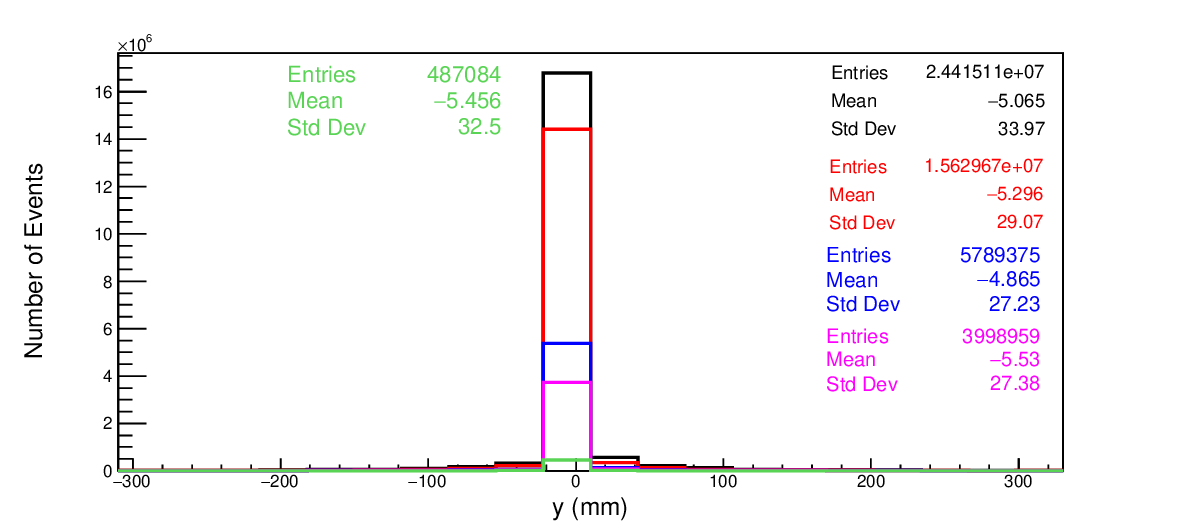}\\
\centering\includegraphics[height=.3\textheight]{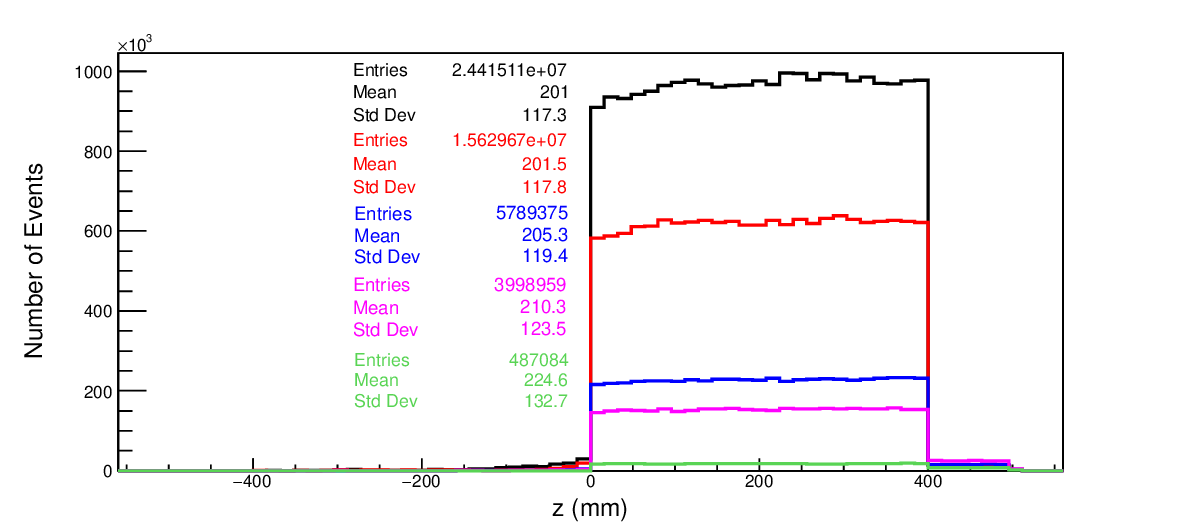}
\caption{The x, y, z distributions of the primaries obtained from simulation of positrons using five different noble gases respectively. Where black, red, blue, pink and green represents xenon, krypton, argon, neon and helium respectively.}\label{figFour}
\end{figure*}

The simulation of electrons and positrons was done to extract the information about the primaries. The electrons or positrons have been simulated in each noble gas one by one and then analyzed. The figure~\ref{figThree} and figure~\ref{figFour} shows the x, y, z distributions of the primaries obtained when 10,000 electrons or positrons of 1 GeV energy were shot from x = 0 cm, y = 0 cm and z = 40 cm position in the gas volume. The different colours in the two figures represents x, y, z distributions of the primaries obtained from simulation of charged particles with five different noble gases. The colouring scheme in the figures shows black, red, blue, pink and green for xenon, krypton, argon, neon and helium respectively. The lower is the ionization energy, the higher is the number of primaries. Therefore xenon gas has highest number of primary ionization as it has lowest ionization energy among other four gases. Also these figures have been obtained after separation of secondaries from primaries. Both figure~\ref{figThree} and figure~\ref{figFour} shows that the x and y distributions have Gaussian shape due to the large number of events. A slight deviation from zero is due to the shooting of particles from x= 0 cm, y = 0 cm and z = 40 cm position. The z distribution shows that electrons were shot along positive z-axis from 40 cm. The same pattern of decrease in primary ionization was observed along z-direction i.e., from xenon to helium as xenon gas has lowest ionization energy. 

Tables~\ref{FirstTable} and ~\ref{SecondTable} show the number of primaries that have been separated from secondaries. The primaries are higher than secondaries due to their general production mechanism. The xenon gas has highest number of ionization as discussed earlier. The conclusions have been discussed in the next section.

\begin{table*}[htb]
\caption{The total number of ionization, primaries and secondaries formed by the interaction of electrons with different noble gases.}\label{FirstTable}
\begin{tabular}{|c|c|c|c|}
  \hline 
  Gases & Total ionization & Primaries (P) & Secondaries (S) \\ \hline\hline
Helium & 772978 & 489574 &  283404 \\ \hline
Neon & 6494247 & 4002013 &  2492234 \\ \hline
Argon & 9470288 &  5791722 &  3678566\\ \hline
Krypton & 25218291 & 15576130 &  9642161 \\ \hline
Xenon & 38312764 & 24290820 & 14021944 \\ 
\hline
\end{tabular}
\end{table*}
\begin{table*}[htb]
\caption{The total number of ionization, primaries and secondaries formed by the interaction of positrons with different noble gases.}\label{SecondTable}
\begin{tabular}{|c|c|c|c|}
  \hline
  Gases & Total ionization & Primaries (P) & Secondaries (S) \\ \hline\hline
Helium & 770179 & 487084 & 283095 \\ \hline
Neon & 6490394 & 3998959 & 2491435 \\ \hline
Argon & 9466898 & 5789375 & 3677523 \\ \hline
Krypton & 25295812 & 15629670 & 9666142 \\ \hline
Xenon & 38497525 & 24415110 &  1408241 \\
\hline
\end{tabular}
\end{table*}

\section{Summary and Conclusions}
The presented studies on the study of simulation of interaction of charged particles with noble gases have been chosen in order to study primary ionization. The noble gases were considered for the study as they have good dielectric properties. In the high energy physics experiments the gaseous detectors in which the particles passes inside the gas medium to produce primary ionization which is required for gaseous detectors. The spatial distributions of primaries after the interaction of electrons and positrons with five different noble gases have been obtained. Also, the number of primaries have been separated from the secondaries. These secondaries further are the source of electrical discharges \cite{citation29}. The discharges cause short-circuit between the electrodes of the detectors which are harmful for the detectors \cite{citation21}. The high energy physics detectors take data for many years and cost of producing the detectors is very high. Any damage in the detector cause loss of data, cost and time which is undesirable. In order to avoid such scenario the simulations of the detectors with different parameters should be done to optimize them. In the end it has been concluded that the secondaries that can cause damage to the detectors needs to be separated that was obtained in the current studies. Also, it has been found that among all the noble gases, the xenon gas is the best one as it produces highest primary ionizations which can be beneficial for the gaseous detectors used in high energy physics experiments.

\section*{Acknowledgement}
I would like to thank the department of physics, Chandigarh University for help and support. I also thank my students for supporting me.

\end{document}